\documentclass[12pt]{iopart}
\usepackage{iopams} 
\newcommand{\be}{\begin{equation}}  
\newcommand{\ee}{\end{equation}}
\newcommand{\bea}{\begin{eqnarray}}
\newcommand{\eea}{\end{eqnarray}}

            \def\l{{}_\lambda}
            
            \def\m{{}_\mu}
            \def\M{{}^\mu}
            \def\n{{}_\nu}
            \def\N{{}^\nu}

\begin{document}
%
%

\title[Particle creation and particle number in an expanding universe]{Particle creation and particle number in an expanding universe}

\author{Leonard Parker}

\address{Physics Department, University of Wisconsin-Milwaukee, Milwaukee, WI 53200, USA}
\ead{leonard@uwm.edu}

\begin{abstract}
I describe the logical basis of the method that I developed in 1962 and 1963 to define a quantum operator corresponding to the observable particle number of a quantized free scalar field in a spatially-flat isotropically expanding (and/or\,contracting) universe.  This work also showed for the first time that particles were created {\em from the vacuum} by the curved space-time of an expanding spatially-flat FLRW universe.  The same process is responsible for creating the nearly scale-invariant spectrum of quantized perturbations of the inflaton scalar field during the inflationary stage of the expansion of the universe.  I explain how the method that I used to obtain the observable particle number operator involved adiabatic invariance of the particle number (hence, the name adiabatic regularization) and the quantum theory of measurement of particle number in an expanding universe.  I also show how I was led in a surprising way, to the discovery in 1964 that there would be no particle creation by these spatially-flat FLRW universes for free fields of any integer or half-integer spin satisfying field equations that are invariant under conformal transformations of the metric. The methods I used to define adiabatic regularization for particle number, were based on generally-covariant concepts like adiabatic invariance and measurement that were fundamental and determined results that were unique to each given adiabatic order.

\end{abstract}

\section{Introduction}
\label{sec-intro}
It is my great pleasure to write this article for a special issue of \JPA in honor of Professor Stuart Dowker. I am particularly grateful to him for his outstanding contributions to quantum field theory in curved space-time, from which I first learned about the beautiful method of zeta-function regularization in curved space-time \cite{Dowker-Critchley1976, Dowker-Critchley1977, Hawking1977, Parker-Cargese1979}.

It seems particularly appropriate for me to describe here the regularization technique that I developed in 1962--1964, while working on my Ph.D. thesis \cite{Parker1966},
to define an operator corresponding to the observable particle number of a free scalar field in a spatially-flat isotropically expanding (and/or\,contracting) universe. This construction appears in detail in \cite{Parker1966}, but until now I have not discussed it fully in the literature. I hope this discussion will clarify the basis and uniqueness of the construction, as well as its intimate relation to particle creation by the gravitational field of the expanding universe. As you will see, the method is {\em not} limited to weak gravitational fields or to any particular form of the change in the scale factor of the universe, $a(t)$, beyond the continuity of $a(t)$ and of a sufficient number of its derivatives.

The need for regularization of the particle number observable became clear to me only after I had considered a quantized field in a smoothly expanding spatially-flat FLRW universe that started from a Minkowski space at early-times and ended in a Minkowski space at late-times. There was no question how to define the vacuum state and particle number of a free quantized field in Minkowski space, and there was no doubt how to propagate the field into the curved space-time of the expanding universe using the generally-covariant equation for the free field obtained by the methods of general relativity. Therefore, I was able to show convincingly that particles were definitely created from the early-time Minkowski vacuum state by showing that after the expansion of the universe there were particles present in the late-time Minkowski space.  The gravitational field of the expanding universe was regarded as a classical field while the quantized field was treated using quantum field theory propagated into the curved space-time of the expanding universe by the generally-covariant free field equation. Perturbations, such as that of the gravitational field (and other classical fields) satisfy linearized field equations, so the same methods also apply to such perturbations, which can be treated as quantized fields that are also created by the expansion of the universe. For a {\em statically bounded} (or asymptotically static) smooth expansion of the spatially-flat FLRW universe, I found that no regularization was needed, apart from the usual Minkowski-space vacuum subtractions at the beginning and end of the expansion, to {\em prove} that in a co-moving volume (1) the particle number was an adiabatic invariant, and (2) that a smooth {\em generic} statically bounded expansion of the universe, starting from the initial {\em Minkowski space vacuum state}, would create particles as a result of the expansion of the universe, giving a non-zero expectation value of the particle number operator in the late-time Minkowski space. The total number density of created particles, summed over all modes of the quantized field, was finite.

Let me digress for a moment to point out that this particle creation is the same process that is responsible for the creation of the nearly scale-invariant power spectrum of primordial inflaton perturbations created by the inflationary stage of the inflationary model of the early universe. A nice proof was given in \cite{Glenz-Parker2009}.  We proved this by joining an asymptotically static smooth expansion of a spatially-flat FLRW universe that approached a Minkowski space at {\em early} times to an {\em arbitrarily long} period of exponential inflation that was finally joined to another asymptotically static smooth expansion of the spatially-flat FLRW universe that approached another Minkowski space at {\em late} times. The joining of the metrics of these three stages of the expansion were continuous to second-order in time derivatives of the expansion scale (as that was sufficient to have no UV divergences in the created number of particles in any co-moving spatial volume). Furthermore, the solutions of the wave equation were required to have appropriate continuity conditions on their time derivatives. (This problem is mathematically analogous to solving a time-independent quantum mechanical scattering problem in one spatial dimension.) The result was calculated {\em analytically} and then evaluated to {\em many} digits of accuracy.  Our result confirmed that if we started from the Minkowski vacuum state of the inflaton perturbation field at early times, then the state {\em evolved} after a relatively small number of e-foldings into the Bunch-Davies vacuum state of the inflationary universe (except for wavelengths that were so large that they were already outside the inflationary Hubble radius when inflation started and remained outside when inflation ended). The result was that the spectrum of inflaton perturbations by the end of 50 or more e-foldings of inflation was scale invariant for the range of wavelengths one could observe in the present universe. Because the expansion of the universe was asymptotically static in the past and the future, there were no UV infinities that survived to late times. There were also no infrared (IR) divergences. The main point I am making here is that the creation of particles (or quanta of a field) by the expansion of the universe is responsible for producing the spectrum of inflaton perturbations in the standard inflationary scenario. {\em Thus, the agreement of current observations of the CMB power spectrum with the inflationary prediction for the primordial spectrum of inflaton perturbations can be interpreted as evidence in favor of the particle creation process we are discussing in the present paper\/}.

Furthermore, if inflation did begin from some kind of initial state that was not the vacuum, then we may be able to {\em observe} some features of the initial state prior to the beginning of inflation, even if inflation lasted for as many as 60 e-foldings or more. This goes back to another topic that appeared in my Ph.D. thesis and early papers, namely, the {\em stimulated} creation of bosons by the expansion of the universe \cite{Parker1966, Parker1969, Parker1971}  Recently, it was shown that this effect may show up in observations of non-Gaussianities that may appear in the bispectrum of CMB anisotropies \cite{Agullo-Parker2011-1, Agullo-Parker2011-2}, as well as in the trispectrum \cite{Agullo-NavarroSalas-Parker2012}.

Let me now return to our discussion of the method of regularization that I developed in 1963 and 1964 to obtain a suitable particle number operator for a free quantized field in an expanding universe.  Encouraged by the results I had obtained for a statically bounded expansion of the universe, I went on to consider universes that were not statically bounded.  After all, observation had shown that we live in a stage during which the universe {\em is} expanding. However, if I tried to use this same time-dependent particle number operator to define the number of particles present at a time $t$ during which the universe was undergoing a generic expansion with non-zero time-derivatives of the scale factor $a(t)$, then I found that the average number of particles (in a given co-moving volume) when summed over all modes would be {\em infinite}, as the result of a UV divergence proportional to the square of $\dot{a}(t)/a(t)$ and to $\ddot{a}(t)/a(t)$.

The particle number observable of a quantized field in the curved space-time of an expanding universe has certain physically-motivated requirements and limitations that guided my reasoning.  The first physically-motivated assumption, underlying the method of regularization, was that the particle number of a free (non-interacting) quantized field in the {\em limit of an infinitely slow and smooth} expansion of the universe should not change with time.  This assumption that the particle number is an {\em adiabatic invariant} is one fundamental requirement of this regularization method, which therefore is known as the method of {\em adiabatic regularization}. (It is clear that requiring adiabatic invariance does {\em not} imply that the particle number operator we obtain will be valid only for scale factors $a(t)$ that change slowly relative to parameters associated with the free quantized field and the gravitational model under consideration.)
A second physically-motivated assumption fundamental to the regularization method was that the regularized particle number observable must have the non-negative integers as its eigenvalues. (A measurement of the particle number present at a given time may be carried out over an interval of time, just as a measurement of the momentum that a particle had at a given time may require an interval of time to complete.)
A third requirement was that during any period when $a(t)$ was constant, the particle number observable would reduce to the particle number observable that I had used in the statically-bounded expansion, and which I already had {\em proved} was an adiabatic invariant for such statically-bounded expansions.  

In later Sections of this paper, I will describe how I found a particle number operator that satisfied these requirements and gave a finite total number of created particles per unit co-moving volume when summed over all modes of the field at any time during the expansion of the universe, even if the expansion was {\em not} statically bounded. In defining the particle number operator, I made the {\em minimal} number of adiabatic subtractions that were necessary to obtain a finite total number of created particles per unit co-moving volume, while satisfying the requirements just described. (Some years later, in 1974, S.A. Fulling and I used this method of adiabatic regularization \cite{Parker-Fulling1974, Fulling-Parker1974} to renormalize the energy-momentum tensor, which required one more term in the adiabatic series for its regularization, as compared with the adiabatic regularization of the particle number.)

In the course of obtaining these adiabatic subtraction terms, I found that there was a special equation of motion of the field for which there was no creation of real particles for {\em any} given smooth expansion of the universe. (By saying {\em real} particles here, I exclude {\em virtual} pairs of particles that would contribute to vacuum energy and pressure during the expansion of the universe, but would not be present at late times in a statically bounded expansion of the universe.) I showed that particles satisfying this special equation of motion had a {\em unique} well-defined vacuum state throughout {\em any} smooth expansion of the universe.  By comparing that special equation of motion with the actual equation of motion of the particles under consideration, I was able to obtain expressions for the subtraction terms that satisfy the requirements stated in the previous paragraph.

I had completed this work in 1963 for the scalar field and during 1964 was in the process of extending these results to spin-1/2 fields, when I happened to see a 1964 article of Roger Penrose \cite{Penrose1964} in which he wrote down the particular wave equation for a scalar field that was invariant under conformal transformations of the metric. I was startled to see that this equation was {\em exactly} the same as the special equation for the quantized scalar field that I had earlier found gave rise to no creation of particles and had a unique well-defined vacuum state throughout any smooth expansion of the spatially-flat FLRW universe.

The remarkable identity of these two equations, led me to find a straightforward proof that there was no particle creation for free fields of {\em any} integer or half-integer spin, that satisfied the {\em conformally invariant} field equations in the spatially-flat FLRW universe. This proof was completed in 1964 and was included in my Ph.D. thesis.  Among the conformally-invariant field equations are the massless Dirac equation and the Maxwell spin-1 photon field equation.  However, one should note that the spin-2 linearized graviton equation obtained from the Einstein gravitational field equations is {\em not} the same as the {\em conformally-invariant} spin-2 field equation.

In the following Sections, I will try to elucidate these introductory remarks relating to the discovery of the creation of particles from the vacuum in an expanding universe and the related development of the method of adiabatic regularization of the particle number observable in such a universe. 

\section{A Brief Orientation}
\label{sec-orient}

In 1962, when I began my thesis work at Harvard under the direction of Sidney Coleman, my objective was to study the possibility that particles of quantized fields in the curved space-time of an isotropically  expanding (FLRW) universe would be created from the vacuum as a result of the expansion (or contraction) of the universe.  At the time, I was not aware of some previous work by Schr\"{o}dinger related to this topic \cite{Schrodinger1939, Schrodinger1940}.  A brief discussion of his work is given near the end of Section \ref{sec-higherspin}.

In curved space-time, I soon realized that methods like normal ordering that set the energy of the vacuum state to $0$ were not acceptable because the energy density of the quantized field must appear in the Einstein gravitational field equations, and if particle creation by the expanding universe does take place, then the energy density of the created particles would {\it not} be expected to remain $0$ as time evolved, even if one started with an initial vacuum state of $0$ energy.  To understand what must occur if there is particle creation, let us use the Heisenberg picture, in which the state vector does not evolve with time, but the quantized field {\it operators} do evolve.

In the Heisenberg picture, suppose that the state vector describing the free quantized field in the expanding universe contains no particles of the field at early times. How may the quantized field evolve with time in such a way that the same state vector at late times will be found to contain a non-zero number of particles of the field?  Such an evolution would require that the annihilation operators of the early-time quantized field evolve into linear combinations of annihilation and creation operators for quanta of the late-time field. (Otherwise, the annihilation operators of the late-time field would annihilate the unchanging state vector, in which case there would be no particles present at late times and the initial vacuum state would also be the final vacuum state). So if particle creation does take place, one must have time evolution under the Heisenberg equation of motion, such that as the quantized field, say $\phi$ evolves in time, the late time annihilation operator $B_{\vec k}$ is a linear combination of the initial annihilation and creation operators $A_{\vec k}$ and $A_{-\vec k}{}^{\dagger}$, 
\[ B_{\vec k} = \alpha_{k}\, A_{\vec k} + \beta_{k}\, A_{-\vec k}{}^{\dagger}.  \]
This transformation corresponds to the creation of particle-antiparticle pairs of total momentum $0$ from the vacuum, as required by conservation of momentum. This linear transformation is an example of a Bogolubov transformation, as was first explicitly pointed out in the context of gravitationally-induced particle creation by \cite{Parker1969}. Analogous Bogolubov transformations that involve a sum over different momenta, as in
\[ B_{\vec k} =\sum_{\vec k'} [\alpha_{k, k'}\, A_{\vec k'} + \beta_{k, k'}\, A_{-\vec k'}{}^{\dagger}]. \] 
occur in other situations that involve gravitationally-induced particle creation. References may be found, for example, in \cite{Birrell-Davies1982} and \cite{Parker-Toms2009}.

\section{Free Scalar Field  in Expanding Universe}
\label{sec-free}
Let us now consider the case of a free quantized scalar field $\phi$ in a smoothly changing spatially flat FLRW universe with line element
\be 
ds^2 = dt^2 - a^2(t)(dx^2 + dy^2 + dz^2), \label{2-1}
\ee
and equation of motion
\be (\Box + m^2)\phi = 0, \label{2-2}\ee
where $\Box= g\M\N \nabla\m \nabla\n$, with $\nabla\l$ denoting the covariant derivative. For now we are considering the so-called {\em minimally-coupled} scalar field having mass $m$.  More generally, one could include a coupling to the Ricci scalar curvature, $R=g\M\N R\m\n$, namely, $(\Box + m^2 + \xi R)\phi = 0$, where $\xi$ is a dimensionless constant. We work in units with $\hbar$ and $c$ each equal to $1$. 

Writing  (\ref{2-2}) with the metric of  (\ref{2-1}), one finds that
\be 
a^{-3} \partial _t(a^3\partial _t\phi ) - a^{-2} \sum _i
\partial^2_i\phi + m^2 \phi = 0. \label{2-3}
\ee
It is convenient to impose periodic boundary conditions in a cube having
sides of coordinate length $L$ and coordinate volume $V = L^3$.  As in
Minkowski spacetime, this is a mathematical device, with $L$ taken to infinity
after physical quantities are calculated.  Then we can expand the field operator $\phi$ in
the form
\be
\phi = \sum _{\vec k}\left\{ A_{\vec k} f_{\vec k}(\vec x, t)
+ A^\dagger_{\vec k} f^*_{\vec k}(\vec x, t)\right\}, \label{2-4a}
\ee
where
\be 
f_{\vec k} = (2V a(t)^3)^{-1/2} e^{i\vec k\cdot\vec x} h_k(t). \label{2-4b}
\ee
Here $k^i = 2\pi n^i/L$ with $n^i$ an integer, $k = \vert\vec k\vert$, and $h_k(t)$ satisfies
\be
 {d^2\over{dt^2}} h_k +  {k^2\over a^{2}} h_k(t) + m^2 h_k(t ) - {3\over 4} \left({\dot a \over a}\right)^2 h_k(t)
  - {3\over 2} {\ddot a \over a} h_k(t) = 0. \label{2-5}
\ee
We infer that
\be 
(f_{\vec k},f_{\vec k'}) = \delta_{\vec k,\vec k'}, \qquad
(f_{\vec k}, f_{\vec k'}^*) = 0, \label{2-6}
\ee
where the quantities on the left-hand-sides of (\ref{2-6}) denote scalar products that are conserved by virtue of (\ref{2-3}).  These scalar products involve integration over a spatial volume at time $t$ in the FLRW universe. (On the conserved scalar product and its consequences, see \cite{Parker-Toms2009}[Section 2.2, (2.48)].) 
The right-hand-sides of (\ref{2-6}) are necessary to insure that if the scale factor $a(t)$ were to smoothly approach a constant, for example, at late times, then the scalar product would agree with that of a free field in the late-time Minkowski space. (Consistency requires (\ref{2-6}) to hold even if $a(t)$ were not to {\em actually} approach a constant in the distant future or for any particular time-interval.) The constant creation and annihilation operators in (\ref{2-4a}) satisfy the commutation relations,
\be 
\left[ A_{\vec k}, A_{\vec k'}\right] = 0, \hspace{2mm} \left[
A^\dagger_{\vec k}, A^\dagger_{\vec k'}\right] = 0 , \hspace{2mm}
{\rm and} \hspace{2mm}
\left[ A_{\vec k}, A^\dagger_{\vec k'}\right] = \delta_{\vec k,\vec k'} . \label{2-6a}\ee
For the relation of these to spin and statistics in an expanding universe, see \cite{Parker-Toms2009}.

In the next Section, I introduce time-dependent creation and annihilation operators as I did in \cite{Parker1966}[for bosons and fermions], \cite{Parker1968}[for bosons and fermions], \cite{Parker1969}[for bosons], and \cite{Parker1971}[for fermions]. I also discuss the adiabatic invariance of particle number, and in a later Section, the role of measurement of particle number in determining the adiabatic subtraction terms. The creation of particles by the expanding universe in the context of quantum field theory was discovered in conjunction with the previous ideas. The connection of particle creation with conformal invariance was also discovered in the course of developing these ideas in my Ph.D thesis. These topics will arise naturally in the following Sections.

\section{The origin of the method of adiabatic regularization}
\label{sec-origin}
In order to help the reader understand the original physical basis for adiabatic regularization, it is necessary for me to explain some crucial features that appear in detail in my Ph.D. thesis \cite{Parker1966}. In the early papers that I published on the results in \cite{Parker1966}, my emphasis was on particle creation by the gravitational field. The method I developed to regularize the time-dependent particle number during the expansion of the universe was only briefly outlined. 

Here, I will emphasize the basic ideas and go into sufficient detail to show how the regularization was based on generally-covariant concepts, such as adiabatic invariance defined in terms of proper (cosmic) time $t$ along the preferred geodesics in an FLRW universe, and measurement of the number of particles present in a co-moving volume at a given cosmic time. If such a measurement is made over a period of time when particles are being created by the expansion of the universe, then there is an irreducible uncertainty in the result of the measurement. This uncertainty allows sufficient latitude in the definition of the particle number operator to allow us to regularize the original number operator so that it remains hermitian, has non-negative integer eigenvalues, and has no UV infinities when summed over all modes of the field. The uniqueness of the resulting particle number observable, defined up to a given adiabatic order, is briefly explained, as is the minimum adiabatic order (in this case second adiabatic order) required to yield a particle number operator that has no UV divergence during the time when particles are being created. If one considers a statically bounded expansion of the universe, then the total number of particles created as a result of the expansion is the same as I had obtained earlier without regularization.

These features are well illustrated in the spatially-flat FLRW universe.  Unless otherwise noted, the term FLRW universe in this paper refers to the class of {\em spatially-flat} FLRW universes with $a(t)$ smoothly changing. 

There are several basic assumptions that guided my reasoning. The first is that the {\em particle number} in a smoothly expanding universe should be an {\em adiabatic invariant.} This means that in the limit of an infinitely slow and smooth expansion (the so-called adiabatic limit), the particle number should remain constant, in the sense that the change in the particle number should approach zero in the adiabatic limit even if the change in the scale factor that occurs over a very long time, i.e., the ratio $a(t_2)/a(t_1)$, is very large.  The adiabatic approximation of Liouville for the harmonic oscillator with a slowly changing frequency gives the following approximation to the solution of (\ref{2-5}):
\be
h_k(t )\sim (\omega_{k}(t))^{-1/2} \exp (\pm i \int^t \omega_{k}(t') \,dt'), 
\label{2-7}
\ee
where 
\be
\omega_{k}(t) = \sqrt{(k/a(t))^2 + m^2} \, .  \label{2-7a}
\ee
This approximation is good in the limit that all time-derivatives of $a(t)$ smoothly approach $0$.\footnote{The Liouville approximation is closely related to the JWKB approximation developed later to approximate solutions of the Schr\"odinger equation for slowly changing scattering potentials, including the possibility of tunneling.} The two solutions in (\ref{2-7}) are linearly independent. Then the general solution of the second-order ordinary differential equation (\ref{2-5}) can be written in the adiabatic approximation as a linear combination
\bea
h_k(t ) &\sim & \alpha_k \, (\omega_{k}(t))^{-1/2} \exp ( - i \int^t \omega_{k}(t') \,dt') \cr
 & & \mbox{}+ \beta_k \, (\omega_{k}(t))^{-1/2} \exp ( + i \int^t \omega_{k}(t') \,dt'),
\label{2-8}
\eea
where $\alpha_k$ and $\beta_k$ are complex constants that must satisfy
\be
|\alpha_k|^2 - |\beta_k|^2 = 1 \label{2-9}
\ee
because of (\ref{2-6}).  In general, $\alpha_k$ and $\beta_k$ will have a time-dependence that is determined by the solution of the ordinary second order differential equation (\ref{2-5}) with $h_k(t)$ written as the sum of two linearly independent solutions in the form
\bea
h_k(t ) & = & \alpha_k(t) \, (\omega_{k}(t))^{-1/2} \exp ( - i \int^t \omega_{k}(t') \,dt') \cr
 & & \mbox{}+ \beta_k(t) \, (\omega_{k}(t))^{-1/2} \exp ( + i \int^t \omega_{k}(t') \,dt'),
\label{2-8b)}
\eea
with
\be
|\alpha_k(t)|^2 - |\beta_k(t)|^2 = 1. \label{2-9b}
\ee
We will assume that $a(t)$ and all its time-derivatives are smooth and well-defined.  

Now the requirement that the particle number in each mode of the field $\phi(\vec{x},t)$ should be an adiabatic invariant is easily met by factoring out the positive and negative frequency adiabatic approximations that appear in (\ref{2-8}) and including the remaining time-dependence of the exact solutions of (\ref{2-5}) in the definition of time-dependent creation and annihilation operators.  Thus, we rewrite (\ref{2-4a}) in the form
\be
\phi(\vec{x},t) = \sum _{\vec k}\left\{ a_{\vec k}(t) g_{\vec k}(\vec x, t)
+ a^\dagger_{\vec k}(t) g^*_{\vec k}(\vec x, t)\right\}, \label{2-10}
\ee
 where
\be 
g_{\vec k}(\vec{x},t) = V^{-1/2} e^{i\vec k\cdot\vec x}
(2a(t){}^3\omega_{k}(t))^{-1/2} \exp ( - i \int^t \omega_{k}(t') \,dt'). \label{2-11}
\ee
 and
 \be
 a_{\vec k}(t) = \alpha_k(t) \, A_{\vec k} + \beta_k(t)^* \, A_{-\vec k}^\dagger. \label{2-12}
 \ee
 As a consequence of (\ref{2-9}) and other conserved quantities (as discussed in \cite{Parker-Toms2009}), it follows that
 \be 
\left[ a_{\vec k}(t), a_{\vec k'}(t) \right] = 0, \hspace{2mm} \left[
a^\dagger_{\vec k}(t), a^\dagger_{\vec k'}(t) \right] = 0 , \hspace{1mm}
{\rm and} 
\left[ a_{\vec k}(t), a^\dagger_{\vec k'}(t) \right] = \delta_{\vec k,\vec k'} , \label{2-13}\ee
Thus, the $a^\dagger_{\vec k}(t)$ and $a_{\vec k}(t)$ satisfy the commutation relations of creation and annihilation operators at all times $t$.  

Can these operators $a^\dagger_{\vec k}(t)$ and $a_{\vec k}(t)$ be interpreted as annihilation and creation operators for real particles present at time $t$? This interpretation is well-supported and consistent at early and late times when one considers a statically bounded (or asymptotically static) expansion of the universe for which $a(t)$ smoothly approaches a constant value $a_1$ at early times and a constant value $a_2$ at late times.  When I began my Ph.D. thesis work in 1962 this was the appropriate case to look at in order to decide if physical particles were created by the expansion of the universe. The reason for having the geometry approach Minkowski spaces at early and late times was that one knew unambiguously how to define real particles in Minkowski-space quantum field theory.  Furthermore, that interpretation agreed with experimental results in Minkowski space.  I was able to prove by 1963 that real particles were indeed created as a result of any typical statically bounded smooth expansion (or contraction) of the universe.  I also showed that the number of particles was an adiabatic invariant, in the sense that the number of particles present at late times was the same as the number that had been present at early times in the limit of an infinitely slow and smooth expansion of the universe \cite{Parker1966, Parker1968, Parker1969, Parker1971}.

In a later Section, we will take up the interpretation of these operators $a^\dagger_{\vec k}(t)$ and $a_{\vec k}(t)$ during a time when $a(t)$ is smoothly changing. As we will see, the particle number in generic cases involves a UV infinity when summed over all modes of the field.  This UV infinity goes away at early and late times for the class of smooth statically bounded expansions of the universe, so no regularization was required to show that particles were created by a statically bounded expansion. However, after showing with no ambiguity that particles are created in such statically bounded expansions, it became necessary to consider the meaning of particles during a time like today, when the universe is actually expanding.

In my thesis, I developed a method based on an asymptotic series expansion about the adiabatic solution of the field equation to deal with the UV infinities that appear in expectation values of the operator  
$a^\dagger_{\vec k}(t) a_{\vec k}(t)$.  The physical basis for the method was a careful consideration of the measurement of the particle number at a given time when $a(t)$ is smoothly changing as a function of time.  This was the original basis for adiabatic regularization. In the context of particle number its development was guided by the natural requirement that the particle number for the free field should be an adiabatic invariant and that the measured particle number should be an integer.  

We have seen in this Section how the adiabatic condition arose in a natural way when considering particles and their creation by the expanding universe; and we will see in a later Section the role played by the adiabatic condition in the construction of a suitable number operator corresponding to the particle number measured during the expansion (or any smooth change) of the universe.  I am often asked  why adiabatic regularization gives covariant results. The reason is that the concepts of adiabatic invariance, and the requirements we discuss later that must be satisfied in a measurement of particle number in an FLRW universe, are independent of coordinate systems, being based on proper time $t$ along the preferred congruence of geodesics assumed to exist in the FLRW universes. To make this clear to the reader, I will include below a fairly detailed discussion of the considerations concerning measurement of the particle number in an expanding universe. These considerations appeared in detail in my thesis, but were only briefly sketched in my early publications on my thesis (\cite{Parker1968}). Therefore, I think they will be novel and of value to the reader.

Before discussing adiabatic regularization of UV infinities in the expectation value of the particle number summed over all modes, we introduce some background and discuss an important relationship discovered in my thesis that involves particle creation and a preferred vacuum state having a symmetry additional to the usual symmetries of the FLRW universes.

\section{Scalar field equation satisfied exactly by a general adiabatic-type solution}
\label{sec-exactadsol} 

The differential equation having {\em exact} solutions of the form
\bea
h_k(t )_0 & = & \alpha_k \, (W_{k}(t))^{-1/2} \exp ( - i \int^t W_{k}(t') \,dt') \cr
 & & \mbox{}+ \beta_k \, (W_{k}(t))^{-1/2} \exp ( + i \int^t W_{k}(t') \,dt'),
\label{3-1}
\eea
with both $\alpha_k$ and $\beta_k$ independent of time, is
\bea
\ddot{h}_k(t )_0 + \left\{W_{k}(t)^{2} - W_{k}(t)^{1/2}[(d^2/dt^2)W_{k}(t)^{-1/2}]\right\}{h}_k(t )_0 = 0.
\label{3-2} 
\eea
We will assume that the function $W_k(t)$ is continuous and has at least two continuous time-derivatives. 

For the minimally-coupled scalar field of (\ref{2-2}), we can rewrite (\ref{2-5}) in the following form \cite{Parker1966}[p.28]:
\bea  \fl \ \ \ \  
\ddot{h}_k(t ) + \left\{ \omega_{k}(t)^{2} - \omega_{k}(t)^{1/2}[(d^2/dt^2) \omega_{k}(t)^{-1/2}]\right\}{h}_k(t ) 
= 2 \omega_{k}(t) S(k, t) h_k(t),
\label{3-3} 
\eea
where $\omega_k (t)$ given in (\ref{2-7a}).  Then, the function $S(k,t)$ for this minimally-coupled scalar field is defined by
\be 
2\omega_k(t) S(k,t) \equiv C_1(k,t) (\dot a(t)/a(t))^2 + C_2(k,t) (\ddot a(t)/a(t)) ,
\label{3-4}
\ee  
with
\be
C_1(k,t) = {k^4 + 3m^2 a(t)^2 k^2 + (3/4) m^4 a(t)^4 \over (k^2 + m^2 a(t)^2)^2 }
\label{3-5a}
\ee
and
\be
C_2(k,t) = {k^2 + (3/2)m^2 a(t)^2 \over k^2 + m^2 a(t)^2 }
\label{3-5b}
\ee

As an aside, one can show that $2 \omega_{k}(t) S(k, t)$ vanishes in the non-relativistic limit that $(m^2 a^2/k^2) \gg 1$ for the particular solution, $a(t) \propto t^{2/3}$, of the classical Einstein equations for a spatially-flat FLRW universe dominated by non-relativistic particles.  In addition, one can show that $2 \omega_{k}(t) S(k, t)$ vanishes in the relativistic limit that $(m^2 a^2/k^2) \ll 1$ for the particular solution, $a(t) \propto t^{1/2}$, of the classical Einstein equations for a spatially-flat FLRW universe dominated by relativistic particles. This was first pointed out in \cite{Parker1966, Parker1968, Parker1969} and means that in those universes there is no creation of these particles of the dominant type.  This is because the adiabatic solution is the {\em exact} solution of the minimally-coupled free field equation in these two cases, which means that the $|\beta_k| = 0$ for them, so there is no particle creation of the dominant type of particle.

Later we will make use of (\ref{3-3}) in arriving at the adiabatic subtractions that regularize the UV divergences that appear in the total particle number summed over all modes when $a(t)$ is not constant. 

But first we use (\ref{3-3}) to show that there is a particular non-minimally coupled scalar field equation having a definite preferred vacuum state over the entire period during which $a(t)$ is smoothly changing. I found this result in 1963 and it led me to the discovery of a surprising connection between particle creation and a symmetry that I only became aware of a year later in 1964.

\section{Scalar field equation giving no creation of massless particles by {\em any} statically bounded smooth expansion of the spatially-flat FLRW universe}
\label{sec-eqnopc}

In the case when $m=0$ in (\ref{3-3}), we find from (\ref{3-4})--(\ref{3-5b}) that
\bea  \fl \ \ \ \ \ \
\ddot{h}_k(t ) + \left\{ \omega_{k}(t)^{2} - \omega_{k}(t)^{1/2}[(d^2/dt^2) \omega_{k}(t)^{-1/2}]\right\}{h}_k(t )
       = (1/6) R(t) h_k(t),
\label{4-1} 
\eea
where $R(t)$ is the Ricci scalar curvature in the spatially-flat FLRW universe,
\be
R(t) = 6\left[ (\dot a(t)/a(t))^2 + (\ddot a(t)/a(t)) \right].
\label{4-2}
\ee

From this we immediately deduce that if we were to start from the massless scalar field equation similar to the minimally coupled one in (\ref{2-2}), but with an additional term involving $(1/6) R(t)$, namely
\be 
(\Box + (1/6) R)\phi = 0, 
\label{4-3}
\ee
which in the spatially-flat FLRW metric is
\be 
a^{-3} \partial _t(a^3\partial _t\phi ) - a^{-2} \sum _i
\partial^2_i\phi + (1/6) R \phi = 0, \label{4-4}
\ee
then the additional $(1/6) R$ on the left-hand side of (\ref{4-4}) would cancel the same term that appeared on the right-hand-side of (\ref{4-1}) when we had started from the minimally-coupled massless scalar field. Therefore, if we were to start from (\ref{4-3}) instead of (\ref{2-2}), then the same steps that had led to (\ref{4-2}) would lead us to the result that
\bea  \fl \ \ \ \ \ \
\ddot{h}_k(t ) + \left\{ \omega_{k}(t)^{2} - \omega_{k}(t)^{1/2}[(d^2/dt^2) \omega_{k}(t)^{-1/2}]\right\}{h}_k(t )
       = 0.
\label{4-5} 
\eea
But this is the same as (\ref{3-2}), the equation that earlier we noted has as its exact general solution
(\ref{3-1}), which in the present case takes the form
\bea
h_k(t )_0 & = & \alpha_k \, (\omega_{k}(t))^{-1/2} \exp ( - i \int^t \omega_{k}(t') \,dt') \cr
 & & \mbox{}+ \beta_k \, (\omega_{k}(t))^{-1/2} \exp ( + i \int^t \omega_{k}(t') \,dt'),
\label{4-6}
\eea
with $\alpha_k$ and $\beta_k$ any chosen complex numbers independent of $t$ and satisfying (\ref{2-9}).

It is now easy to go over the discussion in Section~\ref{sec-free}, but with (\ref{2-2}) replaced by (\ref{4-3}) and (\ref{2-5}) by (\ref{4-5}), to see that if we choose $\alpha_k = 1$ and $\beta_k = 0$ in (\ref{4-6}), then $a_k(t) = A_k$ and $a^\dagger{}_k(t) = A^\dagger{}_k$ for all $t$, which implies that there is no creation of real particles in any smooth statically-bounded expansion. Consistency then implies that there is no creation of real particles satisfying this free, massless, scalar field equation, (\ref{4-3}), for any smooth non-singular expansion $a(t)$, even if there are no periods when $a(t)$ is not changing with time.  Therefore, for the field satisfying (\ref{4-3}) there exists a preferred vacuum state having no real particles at any time during the expansion. This preferred vacuum state is the state annihilated by the operators $A_k$ for all $k$.

The presence of interactions with other particle fields or of self-interactions would, of course, be expected to result in the creation of particles when $a(t)$ is changing. But here we are only considering the effects of the gravitational metric and its derivatives that appear in the generally-covariant free field equation in (\ref{4-3}) . Notice that the gravitational field does not need to be weak for the present considerations to apply. Therefore, one would expect our results to apply in the early universe for scales below the Planck scale.  In particular, for the minimally coupled field equation, (\ref{2-2}), there would be strong creation of minimally-coupled scalar particles by the gravitational field in the early universe. As noted in \cite{Parker1968, Parker1969}, this creation process in the early universe would react back on the gravitational field through the non-linear gravitational field equations. Many studies of this process and its reaction back on the gravitational field have since been carried out. 

\section{The connection with conformal invariance}
\label{sec-conformal}
About a year after finding that there is no creation of particles of the field satisfying (\ref{4-3}) for any smooth expansion of the universe, I happened to read an article by Roger Penrose \cite{Penrose1964}, in which he considered conformal transformations of the metric of the form $g_{\m \n}(\vec x, t) \rightarrow \Omega(\vec x, t)^2 g_{\m \n}(\vec x, t) \equiv \tilde{ g}_{\m \n}(\vec x, t)$, and wrote down a field equation of a scalar field $\phi(\vec x, t)$  that (with multiplication of the scalar field by a suitable power of $\Omega(\vec x, t)$ to define the conformally transformed scalar field $\tilde{\phi}$) remains the same in terms of the conformally transformed metric $\tilde{g}_{\m \n}$ and field $\tilde{\phi}$. What Penrose stated was that under such a conformal transformation of the metric and field the scalar wave equation
\be 
\left(\, \Box + (1/6) R\, \right)\phi = 0, 
\label{5-1}
\ee
implies that $\tilde\phi$ satisfies
\be 
 \left(\tilde\Box + (1/6) \tilde R\, \right)\tilde\phi = 0, 
\label{5-2}
\ee
where $\tilde\Box$ and ${\tilde R}$ correspond to the $\Box$ operator and scalar curvature $R$, but now calculated using the metric ${\tilde g}_{\m \n}$ in place of ${g}_{\m \n}$.

When I saw that the scalar field equation, (\ref{4-3}), that gives exactly no creation of real particles for any smoothly changing spatially flat FLRW universe was the conformally invariant scalar wave equation, I set out to find out why conformal symmetry is responsible for the existence of the preferred vacuum state in this class of FLRW universes. The reason is the global conformal flatness of the the spatially-flat FLRW universes.

The conformal transformation of the metric that takes one from the spatially flat FLRW to Minkowski space is simply given by
\be
g_{\m \n}(\vec x, t) \rightarrow a(t)^{-2} g_{\m \n}(\vec x, t) = \tilde{ g}_{\m \n}(\vec x, t),
\label{5-3}
\ee
followed by defining the new Minkowski-time variable $t_{M} \equiv \int^{t} a(t')^{-2} dt'$. (This Minkowski time variable $t_M$ is the same as the conformal time variable $\eta$.)  It immediately follows that for non-singular smooth $a(t)$ the conformally transformed equation (\ref{5-2}) with time $t_M$ is simply the free field equation in Minkowski space. Hence, because there is no creation of real particles in Minkowski space there is none in spatially-flat FLRW universes having non-singular smooth scale factors $a(t)$.  The positive and negative frequency solutions in Minkowski space, when $t_M$ is written in terms of $t$ and the solutions are multiplied by the appropriate power of the conformal factor that takes the Minkowski space $\phi$ to the FLRW $\tilde{\phi}$, one finds that the exact positive and negative frequency solutions are identical to the positive and negative frequency solutions in (\ref{4-6}). The existence of a global preferred vacuum state in these FLRW universes, thus follows simply from the existence of a preferred vacuum state in Minkowski space.
For field equations having non-zero $m^2$ or that break conformal invariance in other ways, there is indeed particle creation by spatially-flat FLRW universes. For further details and references, see \cite{Parker-Toms2009}[Chapter 2].

\section{Adiabatic Regularization of the particle number operator}
\label{sec-adreg}

Let us return now to the {\em minimally-coupled} field of equation (\ref{2-2}).
When I attempted to use the particle number operators $a^{\dagger}_{\vec k}(t) a_{\vec k}(t)$, with $a_{\vec k}(t)$ defined in (\ref{2-12}), to calculate the {\em total} number of created particles present in a given co-moving volume at time $t$, the following problem arose.  Suppose that the scale factor $a(t)$ is a smooth function, with $[\dot{a}(t)/a(t)]^2$ and/or $\ddot{a}(t)/a(t)$ non-zero. 
I found that the total {\em density} of particles created from the initial Minkowski vacuum state was divergent because the number of particles created in the co-moving volume did not go to $0$ sufficiently rapidly as $k$ approached infinity. 

In the continuum limit of large coordinate length $L$, the expectation value at time $t$, of the number density of particles that were created from the early-time Minkowski vacuum state is
\be
[L a(t)]^{-3}  \Sigma_{\vec k} \langle 0| a^{\dagger}_{\vec k}(t) a_{\vec k}(t) |0 \rangle \propto 
\int^\infty dk k^2 | \beta_k(t) |^2 .
\label{6-1}
\ee
Here, $|0\rangle$ is the state annihilated by the operators $A_{\vec k}$ for all modes ${\vec k}$.  

The spatially-flat FLRW universes that were most commonly considered at the time were the matter-dominated, radiation-dominated, and steady-state de Sitter universes, for which $a(t)$ was proportional, respectively, to $t^{2/3}$, $t^{1/2}$, and $\exp(H t)$. For these and similar expanding universes, dimensionless quantities such as $[\dot{a}(t)/a(t)]^{2n}\, [k/a(t)]^{-2n}$ and $[\ddot{a}(t)/a(t)]^n\, [k/a(t)]^{-2n}$, for positive integers $n$, are roughly of the same order-of-magnitude, and become small for momenta $[k/a(t)]$ that are large with respect to the Hubble parameter, $\dot{a}(t)/a(t)$.

Therefore, I expanded the dimensionless quantity $\beta_k(t)$ in a series of terms having increasing numbers of time derivatives of $a(t)$.  {\em The number of time derivatives of $a(t)$ that appear in a term is called the  adiabatic order of the term} because terms of successively higher adiabatic order approach $0$ as higher powers of the Hubble parameter, and hence faster in the adiabatic limit of an infinitely slow expansion of the universe. 
 
The leading term in the adiabatic expansion of the dimensionless quantity, $| \beta_k(t) |^2$,  turned out to be of second adiabatic order, having dimensionless terms proportional to $[\dot{a}(t)/a(t)]^{2}\, [k/a(t)]^{-2}$ and $[\ddot{a}(t)/a(t)] \,[k/a(t)]^{-2}$. Therefore, the above integral, (\ref{6-1}), diverges linearly as the upper limit of integration approaches infinity. This UV divergence is not present when the scale factor becomes constant in the late-time Minkowski space. It is caused by virtual particles (discussed in the previous Section) that are present only while the scale factor $a(t)$ of the universe is changing. 

To deal with this infinity, I considered the physical requirements that should be satisfied by a number operator corresponding to the number of real particles that would be measured during a time $t$ when $a(t)$ is smoothly changing.  {\em As for any ideal measurement in quantum theory, there must be a corresponding  hermitian operator that has eigenvalues that correspond to the possible results of the measurement.} Because the result of an ideal measurement of particle number present at a given time should be an integer, 
I required:\\ 

{\em (a) that the eigenvalues of the hermitian particle number operator, for each individual mode of the field, is the set of non-negative integers; and\\

(b) that this hermitian particle number operator would reduce to the number operator $a^{\dagger}_{\vec k}(t) a_{\vec k}(t)$ during any periods of time in which the smooth function $a(t)$ became constant (as for example, at early and late times in an asymptotically static expanding universe).}\\

There is an irreducible uncertainty in measuring the particle number present at a given time $t$ in a given co-moving volume $[a(t) L]^3$ in the expanding universe during a period when $a(t)$ is smoothly changing and the derivatives of $a(t)$ are non-zero. For example, suppose one were to prevent particles from escaping or entering through the boundaries of the co-moving volume for a time interval, $\Delta t$, during which one counted the number present. If there were no particle creation by the expansion of the universe in the co-moving volume, then (as a result of the time-energy uncertainty relation) the number counted would have an uncertainty {\em inversely} proportional to the time of measurement, $\Delta t$. On the other hand, if the particles were being created in the co-moving volume by the expansion of the universe during the time $\Delta t$, there would be an additional uncertainty {\em directly} proportional to $\Delta t$ in the number counted. Hence, there would be a minimum or irreducible uncertainty in the number of particles counted by the observer.

As we shall see, the irreducible uncertainty in the measured particle number of a given quantized field, such as the free scalar field, allows us to define a suitable hermitian particle number operator that satisfies requirements (a) and (b) and that has no UV divergence.
  
Within the theory being considered here, although the {\em linearized perturbations} of the gravitational field could be treated as a quantum field satisfying a linear differential equation analogous to that of the quantized scalar field under consideration, the background FLRW universe itself is regarded as classical.
Because of the incompleteness of the semi-classical theory, it is perhaps no longer surprising (as it was to me in 1964, when I did this work in my Ph.D. thesis) that a regularization and renormalization procedure, such as the one to be described below, is necessary to obtain a suitable number operator satisfying requirements (a) and (b) for this {\em free} (non-interacting) scalar field.

I implemented the requirements (a) and (b) by starting from the number operator $a^{\dagger}_{\vec k}(t) a_{\vec k}(t)$ and defining new creation and annihilation operators by adding terms to $a^{\dagger}_{\vec k}(t)$ and to $a_{\vec k}(t)$ that involved $[\dot{a}(t)]^2$ and $\ddot{a}(t)$ in such a way that the new operators satisfied, up to second adiabatic order, the commutation relations for creation and annihilation operators. This implied that up to second adiabatic order, the new hermitian particle number operator formed from these new creation and annihilation operators had eigenvalues consisting of the non-negative integers, thus satisfying the first requirement. Furthermore, the canonical commutation relations for the quantized scalar field and its conjugate momentum were maintained. This was because the definitions of the new creation and annihilation operators resulted simply in a renormalization of the original frequency $\omega$ of the mode ${\vec k}$ of the field to a new frequency, shifted relative to $\omega$ by a term involving $[\dot{a}(t)]^2$ and $\ddot{a}(t)$.  This frequency shift is of second adiabatic order.  

These terms of second adiabatic order can be said to arise from virtual particles because they do not depend on the past behavior of $a(t)$ and they vanish at early and late times in a statically bounded expansion of the universe. These virtual particles have the effect of renormalizing the frequencies of the modes of the field. 

In my Ph.D. thesis I obtained expansions {\em to all adiabatic orders} of the coefficients $\alpha_{k}(t)$ and $\beta_{k}(t)$ in (\ref{2-12}). These coefficients define $a_{\vec k}(t)$ as a linear combination of the {\em constant} annihilation and creation operators $A_{\vec k}$ and $A^{\dagger}_{\vec k}$, respectively.  Using these adiabatic expansions, I found that I could subtract the terms of second adiabatic order from the {\em exact} coefficients $\alpha_{k}(t)$ and $\beta_{k}(t)$ in such a way that the new adiabatically regularized operators replacing $a_{\vec k}(t)$ and $a^{\dagger}_{\vec k}(t)$ satisfied the required commutation relations for annihilation and creation operators. Hence the eigenvalues of the hermitian number operator formed from these adiabatically regularized operators satisfied requirement (a) that the eigenvalues of the hermitian number operator be the non-negative integers. Requirement (b) is also satisfied because the subtractions involve time-derivatives of $a(t)$ and hence vanish for any period during which $a(t)$ is constant. Hence, the adiabatically regularized operators do reduce to the original $a_{\vec k}(t)$ and $a^{\dagger}_{\vec k}(t)$ if $a(t)$ becomes constant for a period of time. I also showed that this could be continued to higher adiabatic order, but because the expression for the particle number had no UV divergence when the second adiabatic order subtractions were made, I felt that subtractions of higher adiabatic order were not necessary when regularizing the particle number.

I further showed that the result for the adiabatic subtractions was {\em unique} to second adiabatic order. 
For example, one could replace $\omega_k(t)$, throughout our considerations, by the quantity $\sqrt{k/a(t)]^2 + m^2 + F(k, \dot{a},\ddot{a},t)}$, where F is a function that vanishes when $\dot{a}$ and $\ddot{a}$ vanish and which never causes that quantity to vanish or become imaginary.  One would obtain the same second adiabatic-order subtractions required for the creation and annihilation operators to satisfy the appropriate commutation relations to second adiabatic order. Similarly, to second adiabatic order, one would obtain the same result for the regularized quantized field and the renormalized frequency \cite{Parker1966}[pp.63--69, Eqs.(80)--(89)].

Here I will write down and then confirm the result I obtained for the renormalized field from the requirements (a) and (b) by using the adiabatic series expansions for $\alpha$ and $\beta$ and showing that the second adiabatic order changes in the creation and annihilation operators had the effect of renormalizing the frequency of each mode of the field.  I will write the end-result for the minimally-coupled scalar field, in terms of the  regularized creation and annihilation operators and the renormalized frequency for each mode.  As mentioned above, this procedure gives a particle number operator that has {\em no UV infinities} when summed over all modes. I will write down the final result below and then use a relatively simple method (also given in \cite{Parker1966}[pp 155--156]) to confirm the result to second adiabatic order.  This method can be extended to higher adiabatic orders if required to renormalize a physical quantity having higher order UV divergences.

To second adiabatic order, I found that the frequency of each mode is shifted from $\omega$ to $\omega - S$, where $S$ was defined in (\ref{3-3}--\ref{3-5b}).  To confirm that the frequency of each mode $\vec{k}$ is shifted, to second adiabatic order, by $S$, let us now return to the minimally-coupled field equation (\ref{2-2}) for a particle of mass $m$. In a spatially-flat FLRW universe, with metric (\ref{2-1}), the field equation takes the form (\ref{2-3}). The functions $h_k(t)$ are defined in terms of the mode functions $f_{\vec k}(\vec x, t)$ in (\ref{2-4b}).  The $h_k(t)$ satisfy (\ref{2-5}), which is written here again for the convenience of the reader:
\be
 {d^2\over{dt^2}} h_k +  {k^2\over a^{2}} h_k(t) + m^2 h_k(t ) - {3\over 4} \left({\dot a \over a}\right)^2 h_k(t)
  - {3\over 2} {\ddot a \over a} h_k(t) = 0. \label{7-1}
\ee
When we were arriving at the equation for which there is no particle creation (the massless conformally invariant equation), we rewrote the above minimally-coupled scalar field equation in the equivalent form of (\ref{3-3}), which we also rewrite here:
\bea  \fl \ \ \ \  
\ddot{h}_k(t ) + \left\{ \omega_{k}(t)^{2} - \omega_{k}(t)^{1/2}[(d^2/dt^2) \omega_{k}(t)^{-1/2}]\right\}{h}_k(t ) 
= 2 \omega_{k}(t) S(k, t) h_k(t),
\label{7-2} 
\eea
where $\omega_k (t)$ is given in (\ref{2-7a}) and the function $S(k,t)$ is defined by
\be 
2\omega_k(t) S(k,t) \equiv C_1(k,t) (\dot a(t)/a(t))^2 + C_2(k,t) (\ddot a(t)/a(t)) ,
\label{7-3}
\ee  
with
\be
C_1(k,t) = {k^4 + 3m^2 a(t)^2 k^2 + (3/4) m^4 a(t)^4 \over (k^2 + m^2 a(t)^2)^2 }
\label{7-4a}
\ee
and
\be
C_2(k,t) = {k^2 + (3/2)m^2 a(t)^2 \over k^2 + m^2 a(t)^2 } \ \ .
\label{7-4b}
\ee
The function $S$ is clearly of second adiabatic order.

It is not difficult to confirm that (\ref{7-2}) can be written to second adiabatic order, as
 \bea   
\ddot{h}_k(t ) + \left\{ W_{k}(t)^{2} - W_{k}(t)^{1/2}[(d^2/dt^2) W_{k}(t)^{-1/2}]\right\}{h}_k(t ) = 0,
\label{7-5} 
\eea
where $W_{k}(t)$ is here defined as
\be
W_{k}(t) = \omega_k (t) - S(k,t).
\label{7-6}
\ee
with $\omega_k (t)$ defined in (\ref{2-7a}) as $\sqrt{[k/a(t)]^2 + m^2}$. 
The reader will recall from (\ref{3-2}) that the exact general solution of this equation is (\ref{3-1}), where 
$\alpha_k$ and $\beta_k$ are complex {\em constants}, with $|\alpha_k|^2 - |\beta_k|^2 = 1$. The frequency of mode $\vec{k}$ to second adiabatic order is thus given by $W_{k}(t)$ of (\ref{7-6}). For this solution having the exact adiabatic form, there is no creation of real particles to second adiabatic order because the coefficients $\alpha_k$ and $\beta_k$ are constants {\em to that order}.  One can continue to iterate this process to arbitrarily high adiabatic order if $a(t)$ has time derivatives to all orders. This does not mean that $|\beta_k|^2$ must actually be $0$, but only that it must have an essential singularity as $a(t)$ smoothly approaches a constant value.  An example of a $|\beta_k|^2$ with that property is the function $\exp[- (k/ a(t))^2 R(t)^{-2}]$, which has an essential singularity as the time-derivatives of $a(t)$ approach $0$. Here $R(t)$ is the Ricci scalar curvature of (\ref{4-2}).
\\

Heuristically, this shift in frequency is analogous to the shift in the frequency of a pendulum that is undergoing simple harmonic oscillations and is suspended in a viscous fluid.  The viscosity of the fluid causes a shift in frequency that is analogous to the shift in frequency of each mode of the field caused by the presence of short-lived virtual particles. Continuing the analogy, if the pendulum bob is suspended from a string, then gradually changing the length of the string will tend to excite the pendulum from its ground state to higher quantum states, with various probabilities. Such excitations are analogous to the creation of particles caused by the expansion of the universe.

In summary, the minimally-coupled quantized scalar field with frequency renormalized to second adiabatic order and the corresponding creation and annihilation operators $a^{\dagger}_{\vec k}(t)$ and $a_{\vec k}(t)$ that give a particle number operator $N_{\vec k}(t) = a^{\dagger}_{\vec k}(t) a_{\vec k}(t)$ that has no UV divergence when summed over all modes, is
\be
\phi(\vec{x},t) = \sum _{\vec k}\left\{ a_{\vec k}(t) g_{\vec k}(\vec x, t)
+ a^\dagger_{\vec k}(t) g^*_{\vec k}(\vec x, t)\right\}, \label{7-7}
\ee
where $g_{\vec k}(\vec{x},t)$ is
\be 
g_{\vec k}(\vec{x},t) = V^{-1/2} e^{i\vec k\cdot\vec x}
(2a(t){}^3 W_k(t))^{-1/2} \exp ( - i \int^t W_{k}(t') \,dt'). \label{7-8}
\ee
 and to second adiabatic order
 \be
W_{k}(t) = \omega_k (t) - S(k,t).
\label{7-9}
\ee
Also to second adiabatic order,
  \be
 a_{\vec k}(t) = \alpha_k(t) \, A_{\vec k} + \beta_k(t)^* \, A_{\vec k}^\dagger, \label{7-10}
 \ee
 with
 \be  
\left[ a_{\vec k}(t), a_{\vec k'}(t) \right] = 0, \hspace{2mm} \left[
a^\dagger_{\vec k}(t), a^\dagger_{\vec k'}(t) \right] = 0 , \hspace{1mm}
{\rm and}\
\left[ a_{\vec k}(t), a^\dagger_{\vec k'}(t) \right] = \delta_{\vec k,\vec k'} , \label{7-11}\ee
Thus, the $a^\dagger_{\vec k}(t)$ and $a_{\vec k}(t)$ satisfy the commutation relations of creation and annihilation operators to second adiabatic order at each time $t$.  

Because $S(t)$ is of second adiabatic order, it vanishes in the adiabatic limit of an infinitely slow continuous and smooth expansion of $a(t)$ (in which each successively higher derivative of $a(t)$ approaches $0$ as one higher power of an adiabatic parameter that approaches $0$). Thus, in the adiabatic limit, the above expressions (\ref{7-7}), (\ref{7-8}),(\ref{7-10}) and (\ref{7-11}) approach (\ref{2-10})--(\ref{2-13}), implying that the number operator $N_{\vec k}(t)=a^{\dagger}_{\vec k} (t) a_{\vec k}(t)$, formed using the operator $a_{\vec k}(t)$ in (\ref{7-10}), is an adiabatic invariant.
What we have gained by working to second adiabatic order is an expression for a particle number operator $N_{\vec k}(t)$ that makes (\ref{6-1}) finite at all times $t$ when $a(t)$ is smoothly changing, and which reduces to the known exact Minkowski space particle number during any time-interval when $a(t)$ has a constant value (as for example, at early and late times for a statically-bounded or asymptotically-static expansion of the universe). As mentioned earlier, one can continue this process of adiabatic regularization to higher adiabatic orders, depending on how many continuous derivatives of $a(t)$ exist.
Generally, we prefer to make the {\em minimum} number of subtractions necessary to renormalize a physical quantity such as the expectation value of the particle number, in order to avoid introducing unnecessary artifacts of regularization into the predicted results of physical measurements.

The irreducible uncertainty inherent in the physical process of measuring the particle number of the minimally-coupled scalar field has been discussed earlier.  That irreducible uncertainty implies that we need not expect there to be a unique {\em precisely} defined particle number operator during an expansion of the universe that is creating particles. Hence, there is no contradiction in defining the particle number operator for that field only to second adiabatic order. In fact, the adiabatic series for $|\beta|^2$ in general is asymptotic, but not necessarily convergent, as explicitly shown in Chapter 2 of \cite{Parker-Toms2009}.

\section{A few words on fermion and boson fields of higher spin}
\label{sec-higherspin}

In the article of Penrose \cite{Penrose1964} that I mentioned earlier, in addition to the conformally-invariant massless scalar field equation, he wrote the field equations using 2-component spinor index notation for massless fields of higher spin in curved space-time.  (The 2-component massless spin-1/2 field equation is the same as the Pauli equation for a massless neutrino and the spin-1 equation is equivalent to the Maxwell equations in vacuum.)  Penrose pointed out that these field equations of massless fermions and bosons, written using 2-component spinor indices, were all invariant under conformal transformations of the metric and fields. In my Ph.D. thesis \cite{Parker1966}, I used those equations, and essentially the same method of proof that I had used for the free, massless, conformally-invariant scalar field equation, to show that the free massless fermions and bosons described by these conformally-invariant field equations were {\em not} created by smoothly expanding spatially-flat FLRW universes. These results held for photons, massless neutrinos, and the other conformally invariant free massless fields satisfying the given conformally-invariant linear field equations of higher spin. In particular, this result also holds for the spin-2 conformally invariant massless quanta (i.e.,{\em conformally-invariant gravitons}) that were described by the spin-2 field equation in \cite{Penrose1964}. 

However, these conformally-invariant gravitons are {\em not} the same as the gravitons that are obtained from the linearized Einstein gravitational field equations.  In 1946, E.M. Lifshitz \cite{Lifshitz1946} had already proved that in the Lifshitz gauge each of the 2 components of this linearized ``Einstein'' graviton field satisfies an equation that is the same as that of the massless {\em minimally-coupled} scalar field equation. I did not know about the 1946 paper of Lifshitz \cite{Lifshitz1946}, so in \cite{Parker1966, Parker1968, Parker1969} I reported on my result that minimally coupled scalar particles would be created by the FLRW expanding universe, and I also reported on my result that free massless fields satisfying conformally-invariant field equations would not be created by these FLRW universes. In particular, I reported that the spin-2 gravitons that satisfied the {\em conformally invariant} massless spin-2 free field equation would not be created in such a universe. Not being aware of Lifshitz graviton field equation obtained from the Einstein gravitational field equations, I did not use my proof for the minimally-coupled scalar field to report that linearized ``Einstein'' gravitons would indeed be created by an expanding FLRW universe. The creation of these linearized Einstein gravitons was studied some time later \cite{Grishchuk1974, Ford-Parker1977}.

As mentioned above, through an argument based on conformal invariance one can show that the free massless spin-1/2 Dirac equation would not give rise to creation of massless neutrinos in an FLRW universe, 
and that the quantized free Maxwell field equation does not give rise to the creation of photons from vacuum in an FLRW universe. Similarly, the presence of free photons (that originated from distant sources such as stars) in intergalactic space would not give rise to {\em stimulated} creation of photon pairs simply as a result of the expansion of an FLRW universe \cite{Parker1966, Parker1968, Parker1969, Parker1972}.

This brings us to some remarks on work done by Schr{\"o}dinger in 1939 and 1940 \cite{Schrodinger1939, Schrodinger1940}. He considered a quantum mechanical wave packet that is propagating in an expanding FLRW universe.  He showed that such a wave packet would {\em increase} in amplitude and would also produce a weaker backward-going wave packet as a result of the expansion of the universe.  He interpreted this process as implying that a scalar particle (which he considered in his actual derivation) or, more importantly, a {\em photon} moving in a given direction had a non-zero probability of inducing the creation of a pair of photons, one moving in the same direction as the original photon (which continues propagating in its original direction) and the other moving in the opposite direction. The frequency of the created photons would be the same as that of the original propagating photon. This process of pair creation is {\em stimulated} by the existence of a particle propagating in a given direction. In the case of photons, he suggested that light propagating from a source, such as a distant star, would create more photons moving in the original direction (as well as the others member of each created pair moving in the opposite direction). At the time when the pair is created,  the created photons would have the same frequency as the original photon, and that frequency would be red-shifted by the expansion of the universe, just as the original photon's frequency is shifted in the FLRW expanding universe. He called this an ``alarming'' phenomenon. 

In the quantum mechanical context in which Schr\"{o}dinger was working, this process is the {\em stimulated} creation of pairs of photons from the presence of existing photons in the expanding FLRW universe. However, as mentioned earlier in this article, I showed that because photons in the FLRW expanding universe satisfy a conformally invariant equation, there is no creation of photon pairs by the expansion, either of the stimulated variety that he was considering, or of the spontaneously created pairs from the vacuum. 

I had been unaware of this work of Schr\"{o}dinger on stimulated pair creation until the early 1970's, when a former colleague of  Schr\"{o}dinger told me about it.  Soon afterward, I wrote a paper pointing out what I said above \cite{Parker1972}.  Thus, Schr\"{o}dinger need not have been alarmed by this process.

Since there seems to be some confusion about the pair creation process from the vacuum, I should point out that the pair creation of scalar particles from {\em vacuum} was first discovered in my Ph.D. thesis \cite{Parker1966}. The method of adiabatic regularization that I have discussed in the present paper was developed to deal with UV infinities that arise in this process of pair creation from the vacuum during the time when the scale factor of the universe, $a(t)$, is actually changing with time. The stimulated creation of pairs is also discussed in \cite{Parker1966, Parker1968, Parker1969}.

\section{Conclusion}

My main purpose in writing this paper is to explain in some detail how the method of adiabatic regularization stemmed from the fundamental physical requirements that must be met by a particle number operator in an expanding FLRW universe. This development was prompted by the discovery \cite{Parker1966} that particle pairs were created from the {\em vacuum} in an FLRW universe with a smoothly changing scale factor $a(t)$.  The number of particles created in any co-moving volume was finite {\em for each mode} of the quantized field.  Furthermore, the total number summed over all modes was finite {\em if} one counted them after the universe had smoothly stopped expanding.

However, at any given time when the universe was actually expanding, there seemed to be a UV divergence in the total number density of created particles summed over all modes of the field, even if $a(t)$ were slowly changing.  As observations show that we do live in an expanding universe, it became necessary to reconsider the particle number operator during times when $a(t)$ is smoothly changing.  As we have seen, adiabatic regularization started from the physically motivated assumptions that the particle number of the free field in each mode (a) should be an adiabatic invariant, analogous to the quantum number of the harmonic oscillator with a smoothly changing frequency, and (b) should have the non-negative integers as its spectrum of eigenvalues. The existence of particle creation caused by the expansion of the universe implied that the measurement of the particle number during the expansion of the universe would have a minimum uncertainty that depended on the time-interval over which the measurement was made and on the rate at which particles were being created by the expansion of the universe during the measurement. This allowed the possibility of defining a satisfactory number operator during the expansion of the universe, up to a given adiabatic order (defined in terms of number of derivatives of $a(t)$, or equivalently, in terms of powers of an adiabatic parameter that goes to $0$ when the expansion parameter approaches a constant). It turns out that the requirements above could be met, while keeping the total number density of particles finite when summed over all modes. That is the main subject of this paper, and is being explained in detail here because it has not been fully discussed in the literature, outside of \cite{Parker1966}.  In the course of the present discussion, I also take the opportunity to discuss the related topics of conformal invariance and higher spin quantized fields.

In view of Professor Stuart Dowker's interest in renormalization in curved space-time and his outstanding contributions, it seems fitting to present these results in this special issue in his honor.


\section*{References}

\begin{thebibliography}{10}

\bibitem{Dowker-Critchley1976} Dowker, S J and Critchley, R 1976 Phys. Rev. {\bf D13}, 3224.
\bibitem{Dowker-Critchley1977} Dowker, S J and Critchley, R 1977 Phys. Rev. {\bf D16}, 3390.
\bibitem{Dowker-Banach1978} Dowker, S J and Banach, R 1978 J. Phys. A: Gen. Phys., {\bf11}, 2255.
\bibitem{Hawking1977} Hawking, S W 1977 Commun. Math. Phys. {\bf 55}, 133.
\bibitem{Parker-Cargese1979} Parker, L 1979 ``Aspects of Quantum Field Theory in Curved Spacetime:
Effective Action and Energy-Momentum Tensor," in {\it{Recent
Developments in Gravitation: Carg\`{e}se 1978\/}}, edited by M. L\'{e}vy and S. Deser (Plenum
Publishing Corp., New York), pp. 219-273.
\bibitem{Parker1966} Parker, L 1966 {\it The Creation of Particles in an Expanding Universe,\/} Harvard University, Cambridge, Massachusetts.
\bibitem{Glenz-Parker2009} Glenz, M M and Parker, L 2009 Phys. Rev. {\bf D 80}, 063534.
\bibitem{Parker-Fulling1974} Parker, L and Fulling, S A 1974 Phys. Rev. {\bf{D9}}, 341.
\bibitem{Fulling-Parker1974} Fulling, S A and Parker, L  1974 Ann. Phys. (N.Y.) {\bf{87}}, 176.
\bibitem{Penrose1964} Penrose, R 1964 in {\it{Relativity, Groups and Topology\/}}, 
edited by C. DeWitt and B. S. DeWitt (Gordon and Breach, New York), 565, 566.
\bibitem{Schrodinger1939} Schr\"{o}dinger, E 1939 Physica {\bf 6}, 899.
\bibitem{Schrodinger1940} Schr\"{o}dinger, E 1940 Physica {\bf 46}, 25.
\bibitem{Birrell-Davies1982} Birrell, N D and Davies, P C W 1982 {\it{Quantum Fields in Curved Space\/}} (Cambridge University Press, London)
\bibitem{Parker-Toms2009} Parker, L and Toms, D J 2009 {\it{Quantum Field Theory in Curved Spacetime: Quantized Fields and Gravity\/}} (Cambridge University Press, London)
\bibitem{Parker1968} Parker, L 1968 Phys. Rev. Lett. {\bf{21}}, 562.
\bibitem{Parker1969} Parker, L 1969 Phys. Rev. {\bf{183}}, 1057.
\bibitem{Parker1971} Parker, L 1971 Phys. Rev. {\bf{D3}}, 346.
\bibitem{Agullo-Parker2011-1} Agullo, I and Parker, L 2011 Phys. Rev. {\bf{D83}}, 063526.
\bibitem{Agullo-Parker2011-2} Agullo, I and Parker, L 2011 Gen. Relativ. Gravit., DOI: 10.1007/s10714-011-1220-8. First Award in the 2011 Essay Competition of the Gravity Research Foundation.
\bibitem{Agullo-NavarroSalas-Parker2012}   Agullo, I, Navarro-Salas, J and Parker, L  2012 "Enhanced local-type inflationary trispectrum from a non-vacuum initial state"  JCAP (accepted for publication).
\bibitem{Lifshitz1946} Lifshitz, E M 1946 Zh. Eksp. Teor. Fiz. {\bf 16}, 587.
\bibitem{Grishchuk1974} Grishchuk, L P 1974 Zh. Eksp. Teor. Fiz. {\bf 67}, 825; Sov. Phys. JETP {\bf 40}, 409 (1975).
\bibitem{Ford-Parker1977} Ford, L H and Parker, L 1977 Phys. Rev. {\bf D16}, 1601.
\bibitem{Parker1972} Parker, L 1972 Phys. Rev. {\bf{D5}}, 2905.

\end{thebibliography}
\end{document}